\documentclass[a4paper,fleqn,usenatbib,useAMS]{mnras}

\usepackage{graphicx}	% Including figure files
\usepackage{amsmath}	% Advanced maths commands
\usepackage{amssymb}	% Extra maths symbols
\usepackage{multicol}        % Multi-column entries in tables
\usepackage{bm}		% Bold maths symbols, including upright Greek
\usepackage{pdflscape}	% Landscape pages

\usepackage[T1]{fontenc}
\usepackage{ae,aecompl}
\usepackage{mathptmx}

\usepackage{hyperref}

\newcommand{\kms}{\,km\,s$^{-1}$} % kilometres per second

\title[Milky-Way-like galaxies in MaNGA]{Are Milky-Way-like galaxies like the Milky Way? A view from SDSS-IV/MaNGA}

\author[S. Zhou]{Shuang Zhou$^{1}$\thanks{Contact e-mail: \href{mailto:Shuang.Zhou@nottingham.ac.uk}{Shuang.Zhou@nottingham.ac.uk}},
Alfonso Arag{\'o}n-Salamanca$^{1}$,
Michael Merrifield$^{1}$,
Brett H. Andrews$^{2}$,
\newauthor
Niv Drory$^{3}$,
Richard R. Lane$^{4}$
\\
$^{1}$School of Physics \& Astronomy, University of Nottingham, University Park, Nottingham, NG7 2RD, UK\\
$^{2}$
PITT PACC, Department of Physics and Astronomy, University of Pittsburgh, Pittsburgh, PA 15260, USA\\
$^{3}$McDonald Observatory, The University of Texas at Austin, 1 University Station, Austin, TX 78712, USA\\
$^{4}$Centro de Investigaci{\'o}n en Astronomía, Universidad Bernardo O'Higgins, Avenida Viel 1497, Santiago, Chile\\}
\date{Last updated ???; in original form ???}

\pubyear{2022}

\begin{document}

\label{firstpage}
\pagerange{\pageref{firstpage}--\pageref{lastpage}}
\maketitle

\begin{abstract}
In this paper, we place the Milky Way (MW) in the context of similar-looking galaxies in terms of their star-formation and chemical evolution histories. We select a sample of 138 Milky-Way analogues (MWAs) from the SDSS-IV/MaNGA survey based on their masses, Hubble types, and bulge-to-total ratios.  To compare their chemical properties to the detailed spatially-resolved information available for the MW, we use a semi-analytic spectral fitting approach, which fits a self-consistent chemical-evolution and star-formation model directly to the MaNGA spectra.  We model the galaxies' inner and outer regions assuming that some of the material lost in stellar winds falls inwards.  We also incorporate chemical enrichment from type II and Ia  supernovae to follow the alpha-element abundance at different metallicities and locations. We find some MWAs where the stellar properties closely reproduce the distribution of age, metallicity, and alpha enhancement at both small and large radii in the MW. In these systems, the match is driven by the longer timescale for star formation in the outer parts, and the inflow of enriched material to the central parts.  However, other MWAs have very different histories.  These divide into two categories: self-similar galaxies where the inner and outer parts evolve identically; and centrally-quenched galaxies where there is very little evidence of late-time central star formation driven by material accreted from the outer regions.  We find that, although selected to be comparable, there are subtle morphological differences between galaxies in these different classes, and that the centrally-quenched galaxies formed their stars systematically earlier.

\end{abstract}

\begin{keywords}
galaxies: fundamental parameters -- galaxies: stellar content --galaxies: formation -- galaxies: evolution
\end{keywords}

%%%%%%%%%%%%%%%%%%%%%%%%%%%%%%%%%%%%%%%%%%%%%%%%%%

\section{Introduction}

Although our location within it can make the large-scale structure of the Milky Way difficult to study, we can resolve its individual stars, giving us access to important clues to its evolution.  We would like to place this knowledge in the broader context of galaxy formation, but unfortunately external galaxies typically provide orthogonal information: we can more readily study the large-scale properties of such systems, but their distances typically prevents us from resolving individual stars.

However, as reviewed by \citet{Bland-Hawthorn2016}, significant progress has now been made in one direction to reconcile these differing views, with a much better understanding developing of the large-scale physical parameters of the Milky Way. Although long recognised as a spiral disk galaxy, even quantities as basic as its exponential scale-length are hard to determine, with values as varied as~2 and~$5\,$kpc reported \citep[e.g.][]{Sackett1997}, but recent investigations have determined the value to be less than $2.5\,$kpc  \citep[e.g.][]{Juric2008,Bovy2013,Reid2014}. Values for other basic parameters have also become better determined.  The total stellar mass of the Milky Way is estimated to be around $6 \times10^{10}{\rm M}_{\odot}$ \citep[e.g.][]{Flynn2006,McMillan2011,Licquia2015}, while its  star formation rate (SFR) is pinned down to $1-2\,{\rm M}_{\odot}{\rm yr}^{-1}$ \citep[e.g.][]{Smith1978,Misiriotis2006,Davies2011,Licquia2015}.

Progress has also continued in studying the properties of the Milky Way's resolved stellar population, and how it varies with location.  It is generally well established that stellar metallicities decline with Galactic radius, but the quality of spectral data means that we can go beyond crude measures of heavy element abundance to look at subtler phenomena such as the prevalence of $\alpha$ elements, where it has been established that, at intermediate metallicities, there seem to be two distinct populations, indicating a complex evolutionary history.  More recent work based on large surveys such as Gaia \citep[e.g.][]{Mikolaitis2014,Recio-Blanco2014}
and SDSS-IV/APOGEE \citep[e.g.][]{Anders2014,Nidever2014,Hayden2015} have mapped out the radial dependence of this subtlety, revealing that stars near the centre of the Galaxy show a strong bimodality in $\rm [\alpha/Fe]$, while stars in the outer disk are more dominated by the low  $\rm [\alpha/Fe]$ population \citep[e.g.][]{Bensby2011,Nidever2014,Hayden2015,Queiroz2020,Griffith2021,Eilers2022}.  

Clearly, any picture of galaxy evolution has to be able to explain such properties. However, before results inferred from the Galaxy can be generalised, we need to have some idea of how typical a galaxy the Milky Way is.  Indeed, there is already some evidence that it is somewhat atypical. 
As pointed out by \cite{Hammer2007}, Milky Way lies out of the Tullly Fisher relation and its halo is made of very low
metal abundance stars. Its small scale-length makes it unusually compact, while its modest star-formation rate and colour place it away from the blue cloud of star-forming galaxies toward the green valley in the colour--magnitude diagram \citep{Schawinski2014}, suggesting that it might be in the process of transitioning to become a quiescent red-sequence galaxy.  

To ascertain how typical the Milky Way is, we really need to define a sample of comparable galaxies, to determine whether there are properties in which it is an outlier.  A number of authors have sought to identify samples of Milky Way analogues \citep[MWAs; e.g.][]{Licquia2016,Kormendy2019,Krishnarao2020,Fraser-McKelvie2019,Boardman2020,Evans2020,Mao2021,Murphy2022}. While there is no unique way to define such a comparison sample, attention has generally focused on similarity in total stellar mass, star-formation rate, and bulge-to-total ratio to select morphologically-similar systems.  For this paper, we identify 138 such MWAs in the SDSS-IV/MaNGA survey \citep{Yana2016}, so that we have high-quality spectra from across the faces of the full sample.

However, even after such a sample has been defined, detailed comparison to the Milky Way remains challenging, because the resolved stellar population data are not available in these external systems.  A range of tools have been developed to investigate stellar populations using the unresolved integrated-light spectra that we have available here, such as {\tt STARLIGHT} \citep{STARLIGHT}, {\tt FIREFLY} \citep{Wilkinson2017} and {\tt pPXF} \citep{Cappellari2004,Cappellari2017}. However, trying to extract both star formation and metallicity information from spectra has proved very challenging, often resulting in inconsistent results and unphysical distributions \citep{Ge2019, Greener2022}. The unrealistic results and limited range of abundance information contained in such fits make it hard to compare them to the Milky Way data.

As a step toward addressing these limitations, we have developed an alternative approach, semi-analytical spectral fitting, which fits in a self-consistent way a full evolutionary model that incorporates both varying star-formation histories and chemical enrichment over time, allowing for both the inflow and outflow of gas \citep[][hereafter Paper~1]{Zhou2022}. In order to investigate the alpha enhancements seen in some Milky Way stars, we have extended the methodology to incorporate the delay in the production of some non-alpha heavy elements by Type Ia supernovae \citep{Worthey1994}.  In addition, since we are interested in radial variations, we have adapted the technique to fit to spectra from multiple spatial zones, whose physics is coupled by the potential transfer of chemically-enriched material between them.  With this new tool, we can reconstruct the star-formation history of different parts of a sample of MWAs in a manner that tracks their metallicity and alpha element abundances, to compare directly with the observed distribution of metallicity and alpha enhancement at different radii in the Milky Way, thus finally ascertaining where our galaxy fits in this bigger picture.

The remainder of this paper is structured as follows.  The Milky Way data, MaNGA data and selection of the 138 MWAs are described in \S\ref{sec:data}, and the modified semi-analytic spectral fitting technique is presented in \S\ref{sec:analysis}. The results comparing the inferred star-formation histories and detailed chemical enrichment of the MWAs to the Milky Way are presented and discussed in \S\ref{sec:results}, with final conclusions drawn in \S\ref{sec:summary}. Throughout this work we use a standard $\Lambda$CDM cosmology with $\Omega_{\Lambda}=0.7$, $\Omega_{\rm M}=0.3$ and $H_0$=70\kms Mpc$^{-1}$.

\section{Data}
\label{sec:data}

\subsection{Milky Way data}
\label{subsec:APOGEE}
There is a growing wealth of observations of the stellar populations of the Milky Way and how they vary with radius. For this analysis, we have adopted data from the Apache Point Observatory Galactic Evolution Experiment (APOGEE) survey \citep{Majewski2017}, which formed part of the Fourth generation of the Sloan Digital Sky Survey, SDSS-IV \citep{Blanton2017}. APOGEE acquired high resolution infrared spectra for more than half a million stars in the Milky Way. Readers are referred to \cite{Holtzman2018} and \cite{Jonsson2020} for further details on the data and data-reduction processes. The data were analysed using the APOGEE Stellar Parameters and
Chemical Abundances Pipeline \citep{GarciaPerez2016}, which provides stellar parameters from the entire spectrum and then individual chemical abundances from windows around the relevant lines.
In this work we use the latest APOGEE data released along with the SDSS-IV Data Release 17 (DR17,\citealt{SDSSDR17}). The ages, distances and chemical abundances that we require for this analysis are taken from the astroNN value-added catalogue (VAC). This VAC applies the astroNN deep-learning code to stellar spectra from APOGEE DR17, from which it determines individual element abundances \citep{Leunga2019} trained with ASPCAP DR17, distances \citep{Leungb2019} trained with Gaia DR3 \citep{GaiaDR32022} and ages \citep{Mackereth2019} trained with APOKASC-2 \citep{Pinsonneault2018} for all the Milky Way stars in SDSS DR17.

\begin{figure*}
    \centering
    \includegraphics[width=0.9\textwidth]{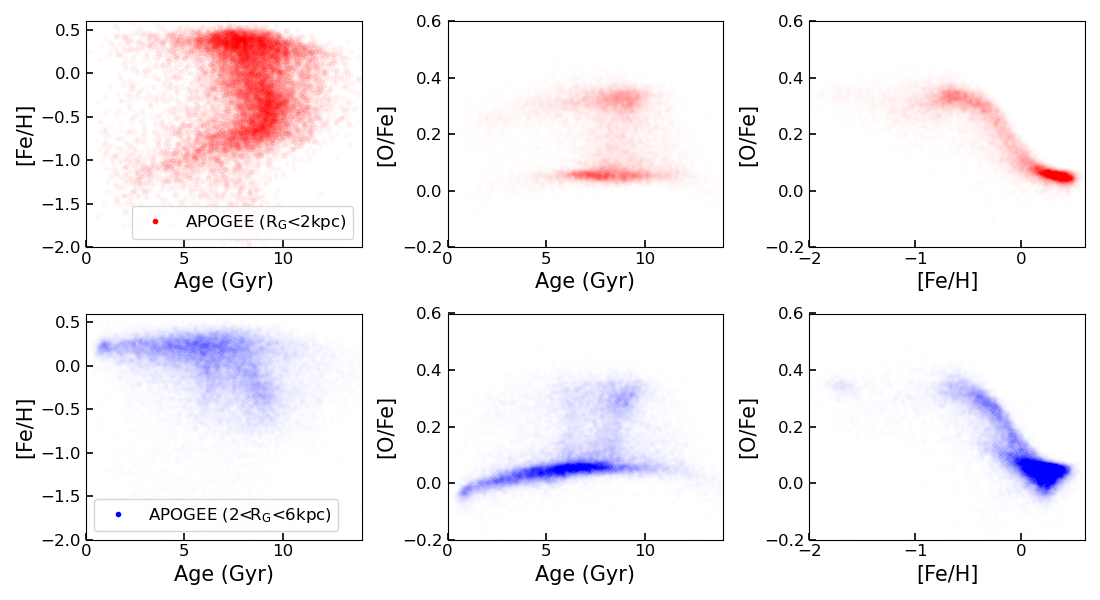}
     \caption{Distributions of Milky Way stars on [Fe/H] vs. age (left), [O/Fe] vs. age (middle) and [Fe/H] vs. [O/Fe] (right) obtained from the APOGEE data.
     Red dots in the first row are stars within $2\,$kpc of the Galactic centre, while blue dots in the second row shows stars in between $2$--$6\,$kpc away from the Galactic centre.
     }
     \label{fig:APOGEE}
\end{figure*}

We want to make a comparison of the radial structure in the Milky Way to external galaxies.  Clearly, we will not have the same wealth of data for external systems, so we look only for the simplest radial effects by dividing the galaxy into an inner and outer region.  In an external galaxy, such a division is most straightforwardly achieved based on the half-light effective radius, $R_{\rm e}$.  In this case, for example, we can conveniently consider an inner region at radii of less than $0.5R_{\rm e}$, and an outer region between 0.5 and $1.5R_{\rm e}$.  For the Milky Way, our location within it means that its effective radius is not that well determined, but if we adopt an exponential scale-length for its disk of $2.7\,$kpc \citep{Licquia2015}, then this component has an effective radius of $4.5\,$kpc.  Adding a modest bulge component brings this figure down somewhat, so we adopt a value of $4\,$kpc. Note that, although adopting a different value for the effective radius may change the relative proportion of high- and low-alpha populations in the central and outer regions of the MW, the overall result that the centre contains two distinct populations (high- and low-alpha stars) while the outer regions are dominated by the low-alpha population, does not depend on the adopted effective radius. In this work we are not making a detailed quantitative comparison since the MWAs selected from MaNGA are not expected to evolve precisely like the MW. Therefore, the exact value adopted for the MW effective radius is not critical for anything that follows. On this basis, we can divide the APOGEE data into inner and outer regions for comparison with the external galaxies, using the cylindrical radii listed in the astroNN VAC.  \autoref{fig:APOGEE} shows the rich structure that exists in the distribution of age and heavy element abundances of stars at small (upper panels) and large (lower panels) radii in the Milky Way. The left panels shows age versus metallicity, the middle shows alpha enhancement versus age, and the right panels show metallicity versus alpha enhancement.  It is apparent from this figure that, after the early formation of alpha-enhanced stars, a metal-rich population began to form in both inner and outer parts of the Milky Way some 10\,Gyr ago, accompanied by a decrease in the level of alpha enhancement, producing the bimodal structure noted in the Introduction.  In the inner parts of the Galaxy the two populations are of similar significance, while in the outer region the low [O/Fe] population is more prominent and is more extended in ages. We cannot hope to reproduce the wealth of structure in this figure for external galaxies where our information is much more limited, nor would we necessarily expect other galaxies to show identical features; the intention is simply to use the overall appearance of this diagram as a template against which other galaxies can be broadly compared, to see whether similar regions of the parameter space are populated, and hence how typical the Milky Way might be.

\subsection{External galaxy data}
For the external comparison data, we have drawn on another of the SDSS-IV projects, Mapping Nearby Galaxies at Apache Point observatory (MaNGA, \citealt{Bundy2015}).  This survey has used integral field units to collect high-quality spectra from across the faces of a representative sample of 10,000 galaxies in the nearby Universe (redshift $0.01<z<0.15$, \citealt{Yana2016,Wake2017}), covering the stellar mass range $5\times10^8 h^{-2}{M}_{\odot} \leq M_*\leq 3 \times 10^{11} 
h^{-2}{M}_{\odot}$ \citep{Wake2017}. Target galaxies are observed out to at least 1.5 effective radii \citep{Law2015}.  The spectra are obtained using the two dual-channel BOSS spectrographs \citep{smee2013} mounted on the 2.5\,m telescope \citep{Gunn2006}, covering $3600-10300${\AA} in wavelength. The data are processed into usable spectra with the bespoke MaNGA Data Reduction Pipeline (DRP; \citealt{Law2016}), and a further passed though the Data Analysis Pipeline (DAP;\citealt{Westfall2019,Belfiore2019}), which provides data products such as the stellar and gas kinematics, as well as emission-line fluxes and spectral indices.
 
\subsection{Milky Way analogue sample selection}
Although the MaNGA survey provides us with a sample of 10,000 galaxies to choose from, selecting precise twins of the Milky Way rapidly reduces the available comparators.  By placing stringent constraints on mass and morphology, \citet{Fraser-McKelvie2019} found that only $\sim0.01\%$ of galaxies match the assumed properties of the Milky Way.  If one places further constraints on star formation rate, the fraction plummets further \citep{Boardman2020}.  Fortunately, though, the intrinsic morphological properties of the Milky Way are not that well determined, so too tight a constraint is not totally justified.  In addition, since we are seeking to place the Milky Way's entire star formation history in the broader context of similar galaxies, we do not want to put too strong a constraint on what a snapshot of its current star formation rate happens to be.  As we will see below, the slightly less restrictive constraints of this project allow us to find a reasonable sample of MWAs.

For the structural parameters of the Milky Way that we are seeking to match, we adopt the values from \citet{Licquia2015}, who found that our galaxy is a relatively massive spiral somewhere in the approximate range $4.94\times10^{10}<M_*/\rm{M}_{\odot}<7.22\times 10^{10}$, with a small bulge that implies a bulge-to-total mass ratio in the range $0.131<\rm(BTR)<0.178$. To identify similar systems in MaNGA, we first use the information contained in the NASA Sloan Atlas catalogue (NSA, \citealt{Blanton2005})\footnote{\label{foot:nsa}\url{ http://www.nsatlas.org/}} to select 
galaxies with stellar mass in the range $4.0\times10^{10}<M_*/\rm{M}_{\odot}<8\times 10^{10}$.  Since the analysis to be undertaken here requires a clear view of any radial variations in the properties of these galaxies, and we need to be able to classify their morphologies fairly unequivocally, we require the systems to be relatively face on, so exclude those with a short-to-long axis ratio $b/a<0.5$ as tabulated in the NSA.

To select spiral galaxies comparable to the Milky Way, we make use of the MaNGA Morphology Deep Learning DR17 Catalogue \citep{Sanchez2022}. This catalogue provides multiple metrics of spiral morphology: 

\begin{enumerate}
    \item the probability of being a late-type galaxy, $P_\text{LTG}$, determined by the Deep Learning process;
    \item an associated estimate of the Hubble T-Type, with $T \leq 0$ for early-type galaxies and $T > 0$ for late-type galaxies;
    \item a visual classification from inspection of the image, in which late-type galaxies have $VC=3$, with a quality flag $VF=0$ for those where this determination is reliable.
\end{enumerate}
 We combine these metrics to narrow the sample to late-type galaxies similar to the Milky Way by selecting systems with $P_\text{LTG} > 0.5$, $T > 0$, $VC = 3$ and $VF = 0$.

Finally, we match the bulge-to-total ratio to the Milky Way, by using the MaNGA PyMorph  photometric catalogue \citep{Sanchez2022}, in which bulges and disks are decomposed by fitting images of galaxies with a Sersic profile bulge component and an exponential disk.  Using these decompositions, we select galaxies in which $0.1 < BTR < 0.2$.  Combining this selection with the morphological and mass cuts, we obtain a final sample of 138 MWAs.

\subsection{Initial processing of the MWA MaNGA data}
\label{subsec:reduction}
MaNGA provides us with many spectra from across the face of each of the sample galaxies, but the typical signal-to-noise ratio of each spectrum is only $SNR \sim 5$ {\AA}$^{-1}$, while the detailed analysis of star formation and chemical evolution requires $SNR \sim 50$.  Fortunately, we are only looking to model the broad properties of the inner and outer parts of each galaxy, which means we can co-add spectra to increase $SNR$.  Accordingly, we divide each galaxy into two regions, an inner part within $0.5R_\text{e}$ (where $R_\text{e}$ is the effective radius as contained in the NSA), and an outer part between $0.5R_{\rm e}$ and $1.5R_{\rm e}$. We co-add the spectra in each region using a similar method to that adopted Paper~1. Briefly, spectra of individual spaxels are first shifted to the rest frame using the velocity measurements available from MaNGA DAP. The fluxes of the spaxels in the two regions are then co-added to obtain their respective stacked spectra, with errors estimated from the inverse variance of the fluxes using standard error propagation. Finally, we correct for inter-spaxel co-variance  using the correction term given by \cite{Westfall2019}. This process results in an inner and an outer spectrum for each galaxy with the requisite $SNR$, typically around 50 per {\AA}.

Our analysis technique also makes use of the current SFR and gas-phase metallicity in the two regions of these galaxies.  Current SFRs are derived from the H${\alpha}$ flux measurements provided by MaNGA DAP. For each pixel, We use the Balmer decrement to correct these fluxes for dust effects, assuming a Calzetti extinction law \citep{Calzetti2000} and an intrinsic H${\alpha}$/H${\beta}$ ratio of 2.87 \citep{Osterbrock2006}. Dust-corrected fluxes from individual pixels are combined to derive total H$\alpha$ in the inner and outer regions.  These fluxes are converted into SFRs using the calibration of \citet{Murphy2011}, assuming a Chabrier initial mass function \citep{Chabrier2003}:
\begin{equation}
\label{eq:sfr}
SFR {\rm (M_{\odot}yr^{-1})=5.37\times10^{-42}}L(\text{H}\alpha).
\end{equation}
To estimate the gas phase metallicity, we adopt the O3N2 method using the calibration of
\citet{PP04}. We obtain measurements of four emission lines, \hbox{[O\,{\sc iii}]}$\lambda$5007, \hbox{[N\,{\sc ii}]}$\lambda$6584, H${\alpha}$, and H${\beta}$, for individual pixels from the MaNGA DAP. The $O3N2$ index for each pixel is then calculated using 
\begin{equation}
 O3N2 \equiv
\rm \log \frac{\hbox{[O\,{\sc iii}]} \lambda 5007/{ H}{\beta}}{\hbox{[N\,{\sc ii}]} \lambda 6584/H{\alpha}},  
\end{equation}
which is converted to an oxygen abundance using
\begin{equation}
12+\log({\rm O/H})=8.73-0.32*O3N2.
\end{equation}

As will discuss in \autoref{subsec:origins}, many of the galaxies discussed in this work have emission line properties similar to low-ionization (nuclear) emission-line regions (LINERs/LIERs). However, the work of \cite{Kumari2019} suggests that the O3N2 metallicity calibration for H{\sc II} regions can also be applied to LIER regions. As we don’t see any significant anomalies in the derived gas-phase metallicities for these galaxies, we use this diagnostic throughout the paper without applying any special treatment to galaxies with possible LIER-like emissions.

As for the SFR, individual pixels are combined in the inner and outer regions to give an estimate of the gas-phase metallicity in each.  Following the approach of paper~1, to allow a direct comparison between this heavy element abundance estimate, the stellar metallicity obtained through population fitting, and the metallicity value in the chemical evolution model, we reconcile the measurements by assuming a solar metallicity of 0.02, and a solar oxygen abundance of $12+\log({\rm O/H})=8.83$ \citep{Anders1989}. Note that there are more recent values of the solar metallicity (0.014) and oxygen abundance ($12+\log{\rm O/H})=8.69$) from \cite{Asplund2009}. However, as we use the \citet{BC03} SSP models with the `Padova1994’ stellar evolutionary tracks, which are  calibrated with the old solar values, we adopt the older value in what follows. This choice will not affect any global trends seen in our results, but appropriate caution is urged when comparing derived  parameters such as  outflow strength with those derived with different calibrations.

\section{Analysis}
\label{sec:analysis}

\subsection{The chemical evolution model}
\label{subsec:model}
We can now turn to comparing the spectral properties of the MWAs to the Milky Way.  As noted, conventional stellar population spectral analysis is not conducive to such a comparison, but the semi-analytic spectral fitting technique does facilitate it.  However, in order to model the more complex data available for the Milky Way, we have to extend the methodology somewhat to allow for radial variations in chemical abundances, and structure in the level of $\alpha$ enhancement seen in stars at different radii and metallicities.  In this section, we describe the revised chemical evolution model and how it is fitted to the MWA spectra.

We first seek to reproduce the evolution of the reservoir of available gas in a galaxy. In Paper~1, this reservoir was treated monolithically as a single entity, with gas flows in due to infalling pristine gas and chemically-polluted outflow in the form of a wind.  However, now we have divided the galaxies into their inner and outer parts, each of which can be thought to have its own reservoir of gas, with the possibility of material being transferred between them.  Indeed, radial inflows have previously included in many of the chemical evolution models that have sought to reproduce the properties of the Milky Way \citep{Mayor1981, Portinari2000, Schonrich2009, Grisoni2018, Palla2020}.  With this addition, the gas mass evolution of the two parts of the galaxy can be described by
\begin{equation}
\label{eq:mass}
\begin{aligned}
&\dot{M}_{\rm g,c}(t)=\dot{M}_{\rm in,c}(t)-\psi_{c}(t)+\dot{M}_{\rm re, c}(t)-\dot{M}_{\rm out, c}(t)+\dot{M}_{r}(t);
\\ 
&\dot{M}_{\rm g,o}(t)=\dot{M}_{\rm in,o}(t)-\psi_{o}(t)+\dot{M}_{\rm re, o}(t)-\dot{M}_{\rm out, o}(t);
\end{aligned}
\end{equation}
The subscript `c' denotes the central/inner region, while `o' denotes the outer region of the galaxy. For each region, the first term at the right-hand side characterises the gas inflow into the region from outside the galaxy, both assumed to have an exponentially decaying form 
\begin{equation}
\begin{aligned}
\dot{M}_{\rm in,c}(t)=A_{c} e^{-(t-t_0)/\tau_c}, \ \ \ t > t_0,
\\
\dot{M}_{\rm in,o}(t)=A_{o} e^{-(t-t_0)/\tau_o}, \ \ \ t > t_0,
\end{aligned}
\end{equation}
where, $t_0$ is the time when the gas starts to infall, with $\tau$ being the infall timescale. As the infall of the gas in a galaxy can be expected to be initiated at approximately the same time across the whole galaxy, we use one single parameter $t_0$ to model the infall time for both regions, but the timescales over which it continues is allowed to differ to model the varying durations of infall expected in different parts of a galaxy, with characteristic timescales $\tau_c$ and $\tau_o$.

The second term in \autoref{eq:mass} represent the gas lost to star formation  and returned by dying stars. Following Paper~1, we adopt a simple Schmidt law \citep{Schmidt1959}, \begin{equation}
\psi(t)=S\times M_{\rm g}(t),
\end{equation}
to model the star formation activity, where the star-formation efficiency $S$ is estimated from the extended Schmidt law \citep{Shi2011} using  
\begin{equation}
\label{eq:sfe}
    S (yr^{-1})=10^{-10.28\pm0.08} \left( \frac{\Sigma_*}{M_{\odot} pc^{-2}} \right)^{0.48},
\end{equation}
The average surface density of stars, $\Sigma_*$, is estimated from the stellar mass maps provided by the Pipe3D catalogue \citep{Sanchez2016}

For the third term in \autoref{eq:mass}, representing the return of gas from stars to the interstellar medium, we assume a constant mass return fraction of $R=0.441$ \citep{Spitoni2017} corresponding to a Chabrier initial mass function, so that the stars formed in each generation will return 44.1\% of their stellar masses to the ISM.

The fourth terms in \autoref{eq:mass} characterise the removal of metal-enriched gas from the two regions. Note that, although the equations are similar, the two terms for the central and outer regions (i.e. $\dot{M}_{\rm out,c}$ and $\dot{M}_{\rm out,o}$) have slightly different physical natures. In the central region the gas removal is only caused by the gas flowing outside the galaxy, while in the outer region the gas removal term also includes the gas
radially transferred into the inner region. However, for simplicity we assume, as is common in chemical evolution models \citep[e.g.][]{Arimoto1987, Spitoni2017}, that its strength is proportional to the star formation activity, with a dimensionless parameter $\lambda$ quantifying its relative strength:
\begin{equation}
\label{eq:outflow}
\begin{aligned}
\dot{M}_{\rm out,c}(t)=\lambda_c\psi_c(t),\\
\dot{M}_{\rm out,o}(t)=\lambda_o\psi_o(t).
\end{aligned}
\end{equation}

The final term in the upper equation of Equations~\ref{eq:mass} characterises the mass growth of the inner region due to the radial flow of processed gas discussed above.  In the interests of capturing the essence of the physics without over-complicating the model, we here make the simple assumption that some fixed fraction $f_r$ of the outflowing gas from the outer region ends up falling into the inner region, so 
\begin{equation}
\label{eq:radialflow}
\begin{aligned}
\dot{M}_{\rm r}(t)=f_r\lambda_o\psi_{o}(t)
\end{aligned}
\end{equation}

Adopting this parameterization, the equations of gas mass evolution can be written as:
\begin{equation}
\label{eq:mass_sum}
\begin{aligned}
&\dot{M}_{\rm g,c}(t)=A_c e^{-(t-t_0)/\tau_c}-S_c(1-R+\lambda_c) M_{\rm g,c}(t)+f_r\lambda_oS_oM_{\rm g,o}(t);\\
&\dot{M}_{\rm g,o}(t)=A_o e^{-(t-t_0)/\tau_o}-S_o(1-R+\lambda_o)M_{\rm g,o}(t)
\end{aligned}
\end{equation}

To follow the chemical evolution of the galaxy, we adopt the usual instantaneous mixing approximation, whereby once metal enriched gas is released by dying stars, it is always well mixed with the ISM. Inflowing gas is assumed to be pristine ($Z=0$), while the outflow will remove metal-enriched gas from the system. Since much of the gas flowing radially to the inner part comes from the mass lost in stellar winds from the outer part, it carries its metallicity with it.  Thus, the equation of chemical evolution can be written as

\begin{equation}
\label{eq:cheevo_c}
\begin{aligned}
\dot{M}_{Z,c}(t)=&-Z_{\rm g,c}(t)(1-R)S_cM_{\rm g,c}(t) + y_Z(1-R)S_cM_{\rm g,c}(t)
\\
&-Z_{\rm g,c}(t)\lambda_c S_cM_{\rm g,c}(t)+f_r\lambda_oS_oM_{\rm g,o}(t)Z_{g,o}(t)
\end{aligned}
\end{equation}
 and 
 \begin{equation}
\label{eq:cheevo_o}
\begin{aligned}
\dot{M}_{Z,o}(t)=&-Z_{\rm g,o}(t)(1-R)S_oM_{\rm g,o}(t) + y_Z(1-R)S_oM_{\rm g,o}(t)
\\
&-Z_{\rm g,o}(t)\lambda_o S_oM_{\rm g,o}(t)
\end{aligned}
\end{equation}
In this equation,  $Z_{\rm g,c}(t)$ and $Z_{\rm g,o}(t)$ are gas phase metallicities in central and outer regions respectively and $M_{Z}(t)\equiv M_{\rm g}\times Z_{\rm g}$. The first term characterises the star formation activity and the heavy elements locked up in long-lived stars, while the second term represents the metal-enriched gas returned by dying stars. The third term describes how the outflow blows away metal-enriched gas, while the final term in \autoref{eq:cheevo_c} represents the metal enrichment arising from radial inflow.

In a comparison to the detailed data from the Milky Way, we need to go beyond considering a single metallicity to look at the abundances of individual elements.  However, there is nothing in the above modelling that does not translate directly into tracing individual elements, as long as they meet the requirement of being returned instantaneously to the interstellar medium.  Thus, for example, we can write equations that dictate the abundances of an $\alpha$ element like oxygen, produced almost entirely in core-collapse supernovae (CCSN) which occur very rapidly, as 
\begin{equation}
\label{eq:cheevo_c_O}
\begin{aligned}
\dot{M}_{O,c}(t)=&-O_{\rm g,c}(t)(1-R)S_cM_{\rm g,c}(t) + y_O(1-R)S_cM_{\rm g,c}(t)
\\
&-O_{\rm g,c}(t)\lambda_c S_cM_{\rm g,c}(t)+f_r\lambda_oS_oM_{\rm g,o}(t)O_{g,o}(t)
\end{aligned}
\end{equation}
 and 
 \begin{equation}
\label{eq:cheevo_o_O}
\begin{aligned}
\dot{M}_{O,o}(t)=&-O_{\rm g,o}(t)(1-R)S_oM_{\rm g,o}(t) + y_O(1-R)S_oM_{\rm g,o}(t)
\\
&-O_{\rm g,o}(t)\lambda_o S_oM_{\rm g,o}(t)
\end{aligned}
\end{equation}
where 
$O_{\rm g,c}(t)$ and $O_{\rm g,o}(t)$ are gas phase oxygen mass fraction in central and outer regions respectively and $M_{O}(t)\equiv M_{\rm g}\times O_{\rm g}$, directly analogous to \autoref{eq:cheevo_c} and \autoref{eq:cheevo_o}.

\begin{table*}
	\centering
	\caption{Priors of model parameters used to fit galaxy spectra}
	\label{tab:paras}
	\begin{tabular}{lccr}
		\hline
		Parameter & Description & Prior range\\
		\hline
		$\tau_c$ & Gas infall timescale in the central region & $[0.0, 14.0]$Gyr\\
		$\tau_o$ & Gas infall timescale in the outer region & $[0.0, 14.0]$Gyr\\
		$t_{0}$ & Start time of gas infall of the galaxy & $[0.0, 14.0]$Gyr\\
		$\lambda_c$ & The wind parameter in the central region & $[0.0, 10.0]$\\
		$\lambda_o$ & The wind parameter in the outer region & $[0.0, 10.0]$\\
		$A_c/A_o$ & Mass gas ratio of gas falling into central and outer regions & $[0.0, 10.0]$\\
		$f_r$ & The radial flow strength & $[0.0, 1.0]$\\
		$E(B-V)_c$&  Dust attenuation parameter in the central region & $[0.0, 0.5]$\\
		$E(B-V)_o$&  Dust attenuation parameter in the outer region  & $[0.0, 0.5]$\\
		\hline
	\end{tabular}
\end{table*}

For other elements such as iron, a significant contribution comes from type Ia supernova (SNIa) \citep[e.g.][]{Nomoto1984,Matteucci2012}, which introduces a delay in the release back into the gas reservoir.  Indeed, this difference from the $\alpha$ elements has proved a key diagnostic in studying the timescales of galaxy formation \cite[e.g.][]{Worthey1994}. SNIa events are thought to arise from the thermonuclear
explosions of white dwarfs in binary stellar systems. However, the details governing physical processes, as well as the progenitors of the SNIa, are still under debate (see reviews by \citealt{Hillebrandt2000} and \citealt{Maoz2014}). For chemical evolution models, especially in modelling the chemical evolution of the Milky Way, the SNIa rates are often modelled by assuming a distribution of time delays before each explosion, but the exact functional form varies in different works \citep[e.g.][]{Kobayashi1998,Greggio2005,Matteucci2009,Snaith2015,Spitoni2019}. 
A more recent investigation by \cite{Gandhi2022} indicates that the SNIa rates can also depend on the metallicity of the stellar populations, which affects the stellar mass-metallicity relations in low-mass galaxies.
The complex physics of SNIa also introduces uncertainties in determining the yields of different elements \citep{Palla2021}. Fortunately, here we are only interested in a broad view of the difference between $\alpha$ and non-$\alpha$ elements, so we can adopt a relatively simple empirical approach. We consider the evolution of iron in the galaxy to be described by two parameters $y_{\rm FeI}$, and $y_{\rm FeII}$. Similar to the total yield of metallicity $y_{\rm Z}$, $y_{\rm FeI}$, and $y_{\rm FeII}$ specify the iron generated in one generation of stars that released by SNIa and CCSN, respectively. The work of \cite{Weinberg2019} found that, to match with the APOGEE observation of the distribution of stars in the [O/Fe] -- [Fe/H] plane, the iron yields from SNIa and CCSN need to be approximately the same. We thus assume 
$y_{\rm FeI}\sim y_{\rm FeII}$. To model the delayed release of half of the iron, we follow \citet{Schonrich2009} by assuming that SNIa are produced in a distribution that begins 0.15\,Gyr after a generation of stars is formed, and their frequency then decays exponentially with an e-folding timescale of 1.5\,Gyr \citep{Forster2006}. Although these assumptions are somewhat arbitrary, we have found through extensive testing that none of the results depend on them at all sensitively. Under these assumptions, the equations that govern the generation of iron can be written
\begin{equation}
\label{eq:cheevo_c_Fe}
\begin{aligned}
\dot{M}_{Fe,c}(t)=&-Fe_{\rm g,c}(t)(1-R)S_cM_{\rm g,c}(t)+y_{\rm{FeII}}(1-R)SM_{\rm g}(t)
\\
&+y_{\rm{FeII}}(1-R)S\int_{t_0}^{t-0.15} M_{\rm g}(t_f)*1.5*e^{-\frac{t-(t_f+0.15)}{1.5}}dt_f
\\
&-Fe_{\rm g,c}(t)\lambda_c S_cM_{\rm Fe,c}(t)+f_r\lambda_oS_oM_{\rm Fe,o}(t)Fe_{g,o}(t)
\end{aligned}
\end{equation}
 and 
 \begin{equation}
\label{eq:cheevo_o_Fe}
\begin{aligned}
\dot{M}_{Fe,o}(t)=&-Fe_{\rm g,o}(t)(1-R)S_oM_{\rm g,o}(t) + y_{\rm{FeII}}(1-R)SM_{\rm g,o}(t)
\\
&+ y_{\rm{FeII}}(1-R)S\int_{t_0}^{t-0.15} M_{\rm g,o}(t_f)*1.5*e^{-\frac{t-(t_f+0.15)}{1.5}}dt_f
\\
&-Fe_{\rm g,o}(t)\lambda_o S_oM_{\rm g,o}(t)\,
\end{aligned}
\end{equation}
where $Fe_{\rm g,c}(t)$ and $Fe_{\rm g,o}(t)$ are the gas phase iron mass fraction in central and outer regions respectively and $M_{Fe}(t)\equiv M_{\rm g}\times Fe_{\rm g}$.

The yields per generation of stars for total metallicity, iron and oxygen can be calculated from individual stellar yields weighted by the initial mass function. There are also plenty of available options in the literature for both CCSN \citep[e.g.][]{Chieffi2004,Nomoto2013} and SNIa \citep[e.g.][]{Iwamoto1999,Thielemann2003,Seitenzahl2013}. \cite{Weinberg2019} investigated many of the yield sets to see whether their predictions can match observations of the Milky Way. Following their work, which matches to the APOGEE data in terms of the evolution of oxygen and iron, we adopt the CCSN yields from \cite{Chieffi2004} and SNIa yields from \cite{Iwamoto1999}. We then calculate  yields by assuming a Chabrier initial mass function \cite{Chabrier2003}, and average over the different metallicities of the progenitors. The resulting mass fractions of oxygen and iron are then converted to [O/Fe] and [Fe/H] using solar values from \citet{Anders1989}.

\subsection{Fitting the model to MaNGA data}
With the model now established, we can turn to how it is fitted to the MaNGA spectroscopic data for the MWAs, for comparison with the Milky Way.  The fit is performed using an updated version of the Bayesian Inference of Galaxy Spectra ({\tt BIGS}), which combine the chemical evolution modelling and sceptical fitting in a Bayesian context. Paper~I has a detailed description of how the approach works and extensive tests of the method; here we summarise the fitting approach, and highlight the refinements needed for this new application.

Initially, we generate a set of model parameters from a proper prior distribution (listed in \autoref{tab:paras}), to calculate the SFHs and ChEHs for both the inner and outer region of the galaxy using \autoref{eq:mass}, \autoref{eq:cheevo_c} and \autoref{eq:cheevo_o}, from which we also obtain the current gas phase metallicities and SFRs for the two regions. We then model the spectra of the two regions assuming these parameter values using the \cite{BC03} stellar population models. Models constructed with the `Padova1994' isochrones, the Chabrier \citep{Chabrier2003} IMF and the STILIB \citep{Borgne2003} stellar templates are used in the analysis. These models cover metallicities from $Z =0.0001$ to $Z = 0.05$, and ages from $0.0001\,{\rm Gyr}$ to $20\,{\rm Gyr}$, and have  spectral resolutions of 3\AA\ FWHM in the wavelength range $3200$ -- $9500$\AA, which is well matched to the spectral range of the MaNGA data, $3600$ -- $10000$\AA, given the median redshift of $z\sim0.03$ of the target galaxies \citep{Wake2017}. To account for the broadening of the observed spectra due to stellar velocity dispersion and instrumental effects, we fit the stacked spectra with the software {\tt pPXF} \citep{Cappellari2017} to obtain an effective velocity dispersion, which is then used to broaden the SSP models and match the spectral resolution. We then compare the spectra, gas phase metallicities and SFRs for both the inner and outer regions of this model with the observed data using a $\chi^2$-like likelihood function,
\begin{equation}
\label{likelyhood}
\begin{aligned}
\ln {L(\theta)}\propto &-\sum_{i}^N\frac{\left(f_{\theta,c,i}-f_{\rm D,c,i}\right)^2}{2f_{\rm err,c,i}^2}-
\frac{(Z_{\rm g,c,\theta}-Z_{\rm g,c,D})^2}{2\sigma_{Z}^2
}-\frac{(\psi_{c,\theta}-\psi_{\rm c,D})^2}{2\sigma_{\psi,c}^2}\\
&-\sum_{i}^N\frac{\left(f_{\theta,o,i}-f_{\rm D,o,i}\right)^2}{2f_{\rm err,o,i}^2}-
\frac{(Z_{\rm g,o,\theta}-Z_{\rm g,o,D})^2}{2\sigma_{Z}^2
}-\frac{(\psi_{o,\theta}-\psi_{\rm o,D})^2}{2\sigma_{\psi,o}^2},\,
\end{aligned}
\end{equation}
where the quantity with subscript `c' and `o' are for the central and outer regions respectively. Quantities with subscript $\theta$ denote model predictions from the parameter set $\theta$, while the subscript D stands for observed data. $f_i$ is the flux of each spectrum at the $i$-th wavelength point, which are compared over al $N$ wavelength points. $Z_{\rm g}$ is the current gas phase metallicity, while $\psi_{0}$ is the current SFR of the region, whose observed values, estimated in \autoref{subsec:reduction}, are compared to those of the model in each region.

The parameters are then varied, as described in Paper~1, to obtain the best-fit model parameters.  From this model, we can extract the full SFHs and ChEHs, but also we can obtain the evolution of iron and oxygen separately, as described above.  This information allows us to see for each MWA how the best-fit model predicts that the age -- [Fe/H] -- [O/Fe] space should be populated by stars, for direct comparison with the APOGEE data from the Milky Way, to finally ascertain how typical our galaxy is.

\section{Results and discussion}
\label{sec:results}
We applied the semi-analytic spectral fitting approach described above to the sample of 138 MWAs selected from MaNGA.  As we will see below, a good fraction have properties that are similar to the Milky Way, but there are some where the detailed chemical properties predicted by the best fit are starkly different.  We will address the possible origins of those differences in \autoref{subsec:origins}.

\subsection{Milky-Way-like MWAs}
\begin{figure*}
    \centering
    \includegraphics[width=0.9\textwidth]{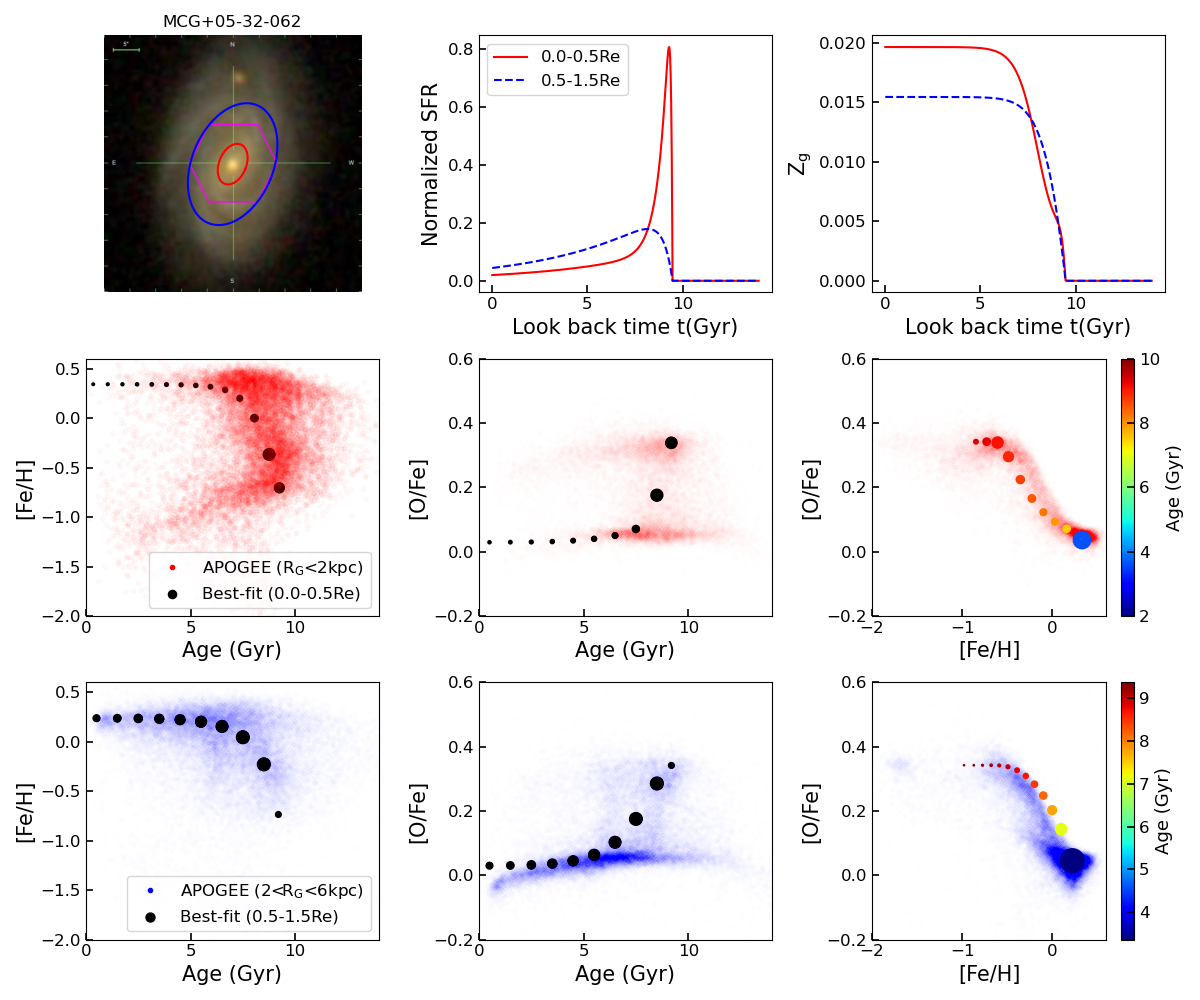}
     \caption{Comparison between our best-fit model of MWA galaxies and APOGEE observation of the Milky Way. In the first row, the first panel shows the optical image of the example galaxy, with the MaNGA footprint shown in magenta. The two ellipses mark the positions of $0.5R_{\rm e}$ (red) and $1.5R_{\rm e}$  (blue) of the galaxy.
     The second panel show the normalized SFH obtained from our best-fit models to the data stacked within $0.0$--$0.5R_{\rm e}$ (red solid) and $0.5$--$1.5R_{\rm e}$ (blue dash), respectively, while the third panel shows the corresponding ChEHs. 
     The second row compares our model predictions of the evolution in $0.0$--$0.5R_{\rm e}$ with the APOGEE measurements of stars within $2\,$kpc of the Galactic centre. The first and  second panel compare the time evolution of iron and oxygen respectively, while the third panel plot the [O/Fe] vs. [FeH]. In each panel, small dots correspond to individual APOGEE stars, while circles are predictions of our model with their sizes indicating the relative fraction of the corresponding stars in the model stellar populations. In the right panel, the colour of the circles marks the ages of the stellar populations. The third row is similar to the second row, but shows the comparison between our model predictions in the outer region ($0.5$--1.5$R_{\rm e}$) and APOGEE stars $2$--$6\,$kpc away from the Galactic centre. 
     }
     \label{fig:example_MWA}
\end{figure*}

In \autoref{fig:example_MWA}, we present the fits to one of the MWAs, MCG+05-32-062 (or 8983-3703 in its MaNGA designation), superimposed on the equivalent APOGEE data for the Milky Way.  In the first row of \autoref{fig:example_MWA} we show the SFH and ChEH of the galaxy derived from the best-fit model, from which we see clearly different evolution between the central and outer regions. The central part of the galaxy (red line), which contains the bulge, formed a significant fraction of its stellar mass in a strong burst of star formation around 10\,Gyr ago, but then continues at a low level to the present day. In contrast, the outer part of the galaxy has a much more extended star formation history. The metallicities in both regions grow quickly in the first few billion years of star formation, but then asymptote to their current values. Looking at the physical parameters in the model that drove this evolution, we find that the inner region has a short gas infall timescale ($\tau_c\sim 0.5$Gyr), which triggers the strong initial burst of star formation. In contrast, the outer region has a much longer gas infall timescale  ($\tau_c\sim 3$Gyr), leading to the more extended SFH. In addition, there is a significant radial inflow in the galaxy, which provide the material for the modest on-going star formation in the central region.
 We model the radial flow as a constant mass fraction relative to the strength of the outflow, while the outflow itself is proportional to the star formation rate (see equations in \autoref{sec:analysis}).  Consequently, the radial flow takes a constant fraction of the gas from the outer region into the inner region in a given time period. For this specific galaxy, the best-fit parameters imply that the radial flow moves materials at a rate of approximately 0.2$M_{o}$/Gyr, i.e., around 20\% of the gas in the outer region will flow into the inner region in 1Gyr. If we simply assume that the gas surface density within 
$1.5R_\text{e}$ of the galaxy is approximately constant, we estimate that for a galaxy with $R_\text{e}=4\;$kpc, this radial flow would have an average velocity of $1.2\,$km/s at the boundary between the two regions ($2\,$kpc radius). This value, albeit rather rough, is consistent with commonly adopted values in chemical evolution models \citep[e.g.][]{Lacey1985,Portinari2000,Spitoni2013}, and in a recent hydro simulation of a MWA \citep{Vincenzo2020}. Moreover, the metal-enriched gas inflow results in the relatively high metallicity in the centre and builds up the metallicity gradient in the galaxy. 

The lower two rows show the comparison to the inner and outer parts of the Milky Way, with the tracks in this parameter space as predicted by the best-fit chemical model from \autoref{subsec:model} superimposed on the APOGEE data from \autoref{subsec:APOGEE}.  The size of the points in the external galaxy model is mapped from the SFR to give an indication of how well populated each part of the tracks should be. Bear in mind that this is not a fit to the Milky Way, but simply our chemical evolution model fitted to the spectra of a galaxy chosen just because it looks superficially rather like the Milky Way.  As such, although not reproducing the wealth of detail in the Milky Way data, it is surprising how well the two trace each other across this range of parameters, even down to details such as the bimodality in the [O/Fe] -- [Fe/H] plane at small radii and its diminution at large radii.  It is clear that there are other galaxies with similar properties to the Milky Way, which can be broadly reproduced by a fairly simple chemical evolution model.

Because the stellar populations are derived from a full chemical evolution model, we can interpret their distribution in MCG+05-32-062, and by association those in the Milky Way, in physical terms.  The significant old, metal-poor but $\alpha$ enhanced population in the centre mainly formed in a strong initial burst triggered by the short timescale gas infall there. These old stars generally have enhanced [O/Fe], as SNIa have not been able to eject much iron for recycling through the ISM on such a short timescale. After this short period, the gas infall to the centre declines and central star formation starts to be fueled by the radial inflow of gas from the outer part. This radial flow is metal enriched but not enhanced in $\alpha$ elements, producing the younger population with solar chemical compositions in the centre. For the outer region, only a small fraction of stars formed early in the galaxy's formation are $\alpha$-enhanced. The majority of stars here formed at a more leisurely rate during the longer-timescale gas infall, during which the ISM has been polluted by SNIa explosions, so that the $\alpha$ abundance is also close to the solar value. 

While we have picked out the example of MCG+05-32-062 for its similarity to the Milky Way, it is far from unique. 56 of the 138 MWAs show a similar pattern of a shorter timescale for star formation at small radii compared to large, followed by lower-level ongoing central star formation fueled by recycled gas from the outer parts. The inflow velocities span a wide range, from $0.3\,$km/s to $2\,$km/s, with a median of $0.8\,$km/s.  It is therefore  clear that the Milky Way's chemical properties and implied formation history are not that unusual. However, there are other galaxies that meet the morphological criteria to be MWAs where the stellar properties inferred from their spectra, and hence the implied evolutionary histories, are rather different.  We turn now to look at these cases. 
\begin{figure*}
    \centering
    \includegraphics[width=0.72\textwidth]{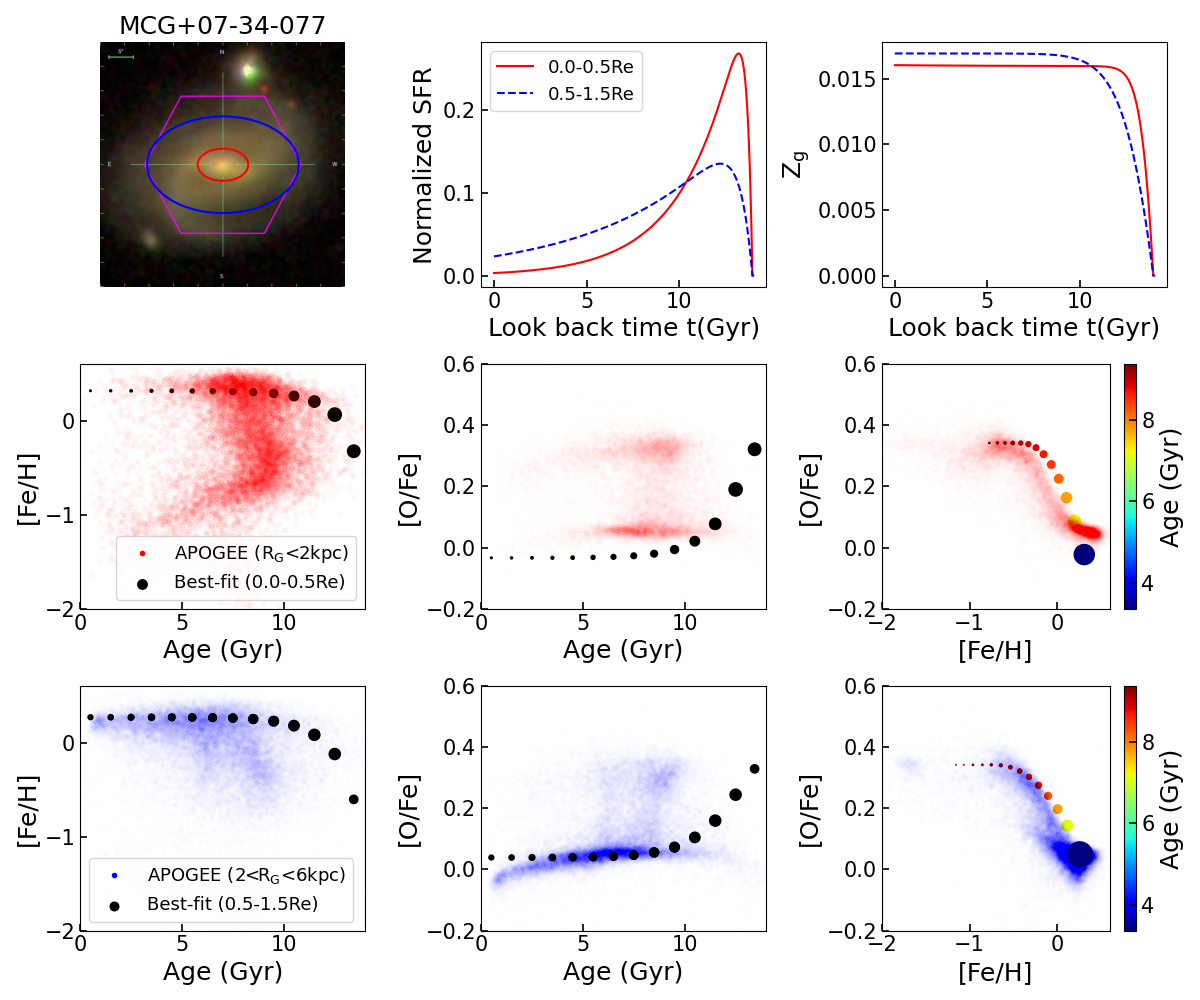}
     \caption{As for \autoref{fig:example_MWA} but for a self-similar galaxy. Note the difference in inferred stellar properties from the Milky Way.
     }
     \label{fig:example_ss}
\end{figure*}
\begin{figure*}
    \centering
    \includegraphics[width=0.72\textwidth]{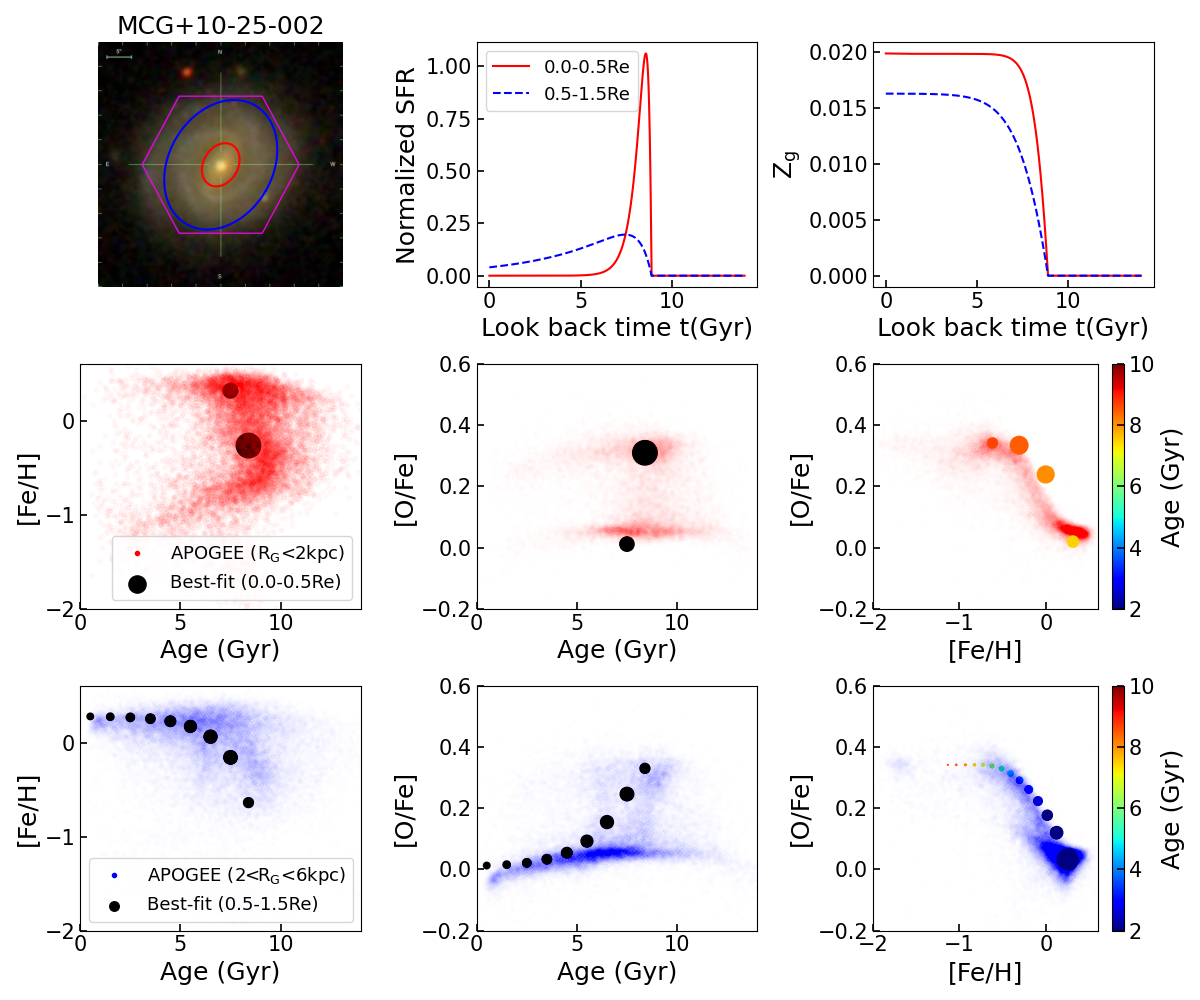}
     \caption{As for \autoref{fig:example_MWA} but for a centrally-quenched galaxy. Note the difference in inferred stellar properties from the Milky Way..
     }
     \label{fig:example_cq}
\end{figure*}

\subsection{MWAs unlike the Milky Way}
Analysis of the spectral properties of the full sample of MWAs reveals two distinct classes of object with histories unlike the Milky Way. Typical examples of \autoref{fig:example_ss} and \autoref{fig:example_cq}.

\autoref{fig:example_ss} shows an object in which star formation in the inner and outer parts of the galaxy are both extended, leading to a similar chemical enhancement at all radii.  Given this similarity, it is unsurprising that the inferred distribution of stars in the age -- metallicity -- alpha enhancement space is very similar at small and large radii, and very unlike the Milky Way. The inferred timescales for gas infall in these ``self similar'' galaxies are comparable to the Milky Way's at a few billion years; the difference arises because there is no variation in timescale with radius. These objects make up 55 out of 138 MWAs, a very similar fraction to those that follow a Milky-Way-like evolution.

The second class of non-Milky-Way-like system is illustrated by \autoref{fig:example_cq}.  In this kind of centrally-quenched galaxy, star formation continues for an extended period in the outer parts much like the Milky Way, but there is no indication in the spectra of recent metal-enriched star formation in the inner parts.  From the model, it is apparent that this phenomenon is driven by an absence of radial gas inflow, with the fitting process strongly preferring to set the associated parameter, $f_r$, at its lower bound of zero.  One concern with such a piling up of parameters at the lowest allowed value is that it might indicate that the model does not explore the parameters sufficiently fully.  However, we have refitted these spectra with a model that considered the opposite possibility of a net flow of processed gas from the inner parts of the galaxy to the outer parts, but in almost no cases did this result in a significantly non-zero value either.  It would appear that for the 27 MWAs in this class, there is no indication in the spectra of any coupling between the evolution of the inner and outer parts of the system driven by wind effects. 

\begin{figure}
    \centering
    \includegraphics[width=0.45\textwidth]{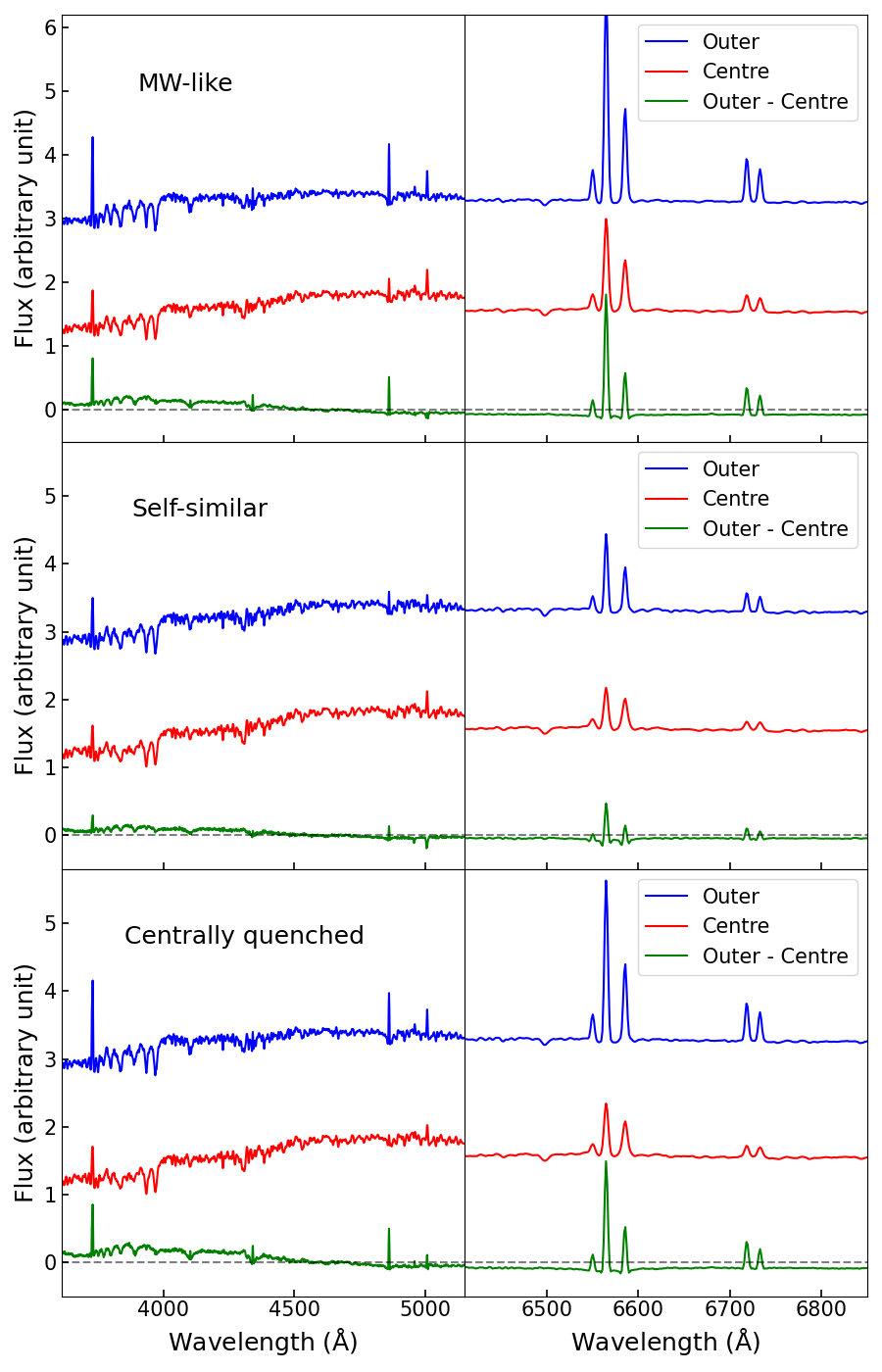}
     \caption{Stacked spectra for Milky-Way-like (top), self-similar (middle) and centrally quenched (bottom) galaxies. The left panels zoom in on the spectral features around 4000{\,\AA}, while the right panels focus on the H$\alpha$ emission line. In each panel, the blue line shows the average spectra of the outer regions, while the red line corresponds to the inner regions. The green lines display the difference between the spectra of the inner and outer regions.
     }
     \label{fig:spec_stack_diff}
\end{figure}

These three separate classes -- Milky-Way-like, self-similar, and centrally quenched galaxies -- not only have best-fit models with very different properties, but also exhibit significant variations in their spectra. In \autoref{fig:spec_stack_diff} we show the average spectra for the three galaxy classes (Milky-Way-like, top panel; self similar, middle panel, and centrally quenched, bottom panes). Milke-Way-like galaxies show strong emission lines throughout, with relatively small but significant differences between the spectra of their central (red) and outer (blue) regions.  In contrast, self-similar galaxies have very similar spectra at all radii, consistent with the self-similar evolution shown by the models in \autoref{fig:example_ss}. The inner regions of the centrally quenched galaxies show very low star formation, as indicated by the weak emission lines, and exhibit the largest spectral differences from the centre to the outer regions. This is particularly evident in the spectral features around 4000{\,\AA} and the Balmer series. These spectra difference give us confidence that our model are able to identify real intrinsic variations in the evolution of these galaxies.

\begin{figure}
    \centering
    \includegraphics[width=0.5\textwidth]{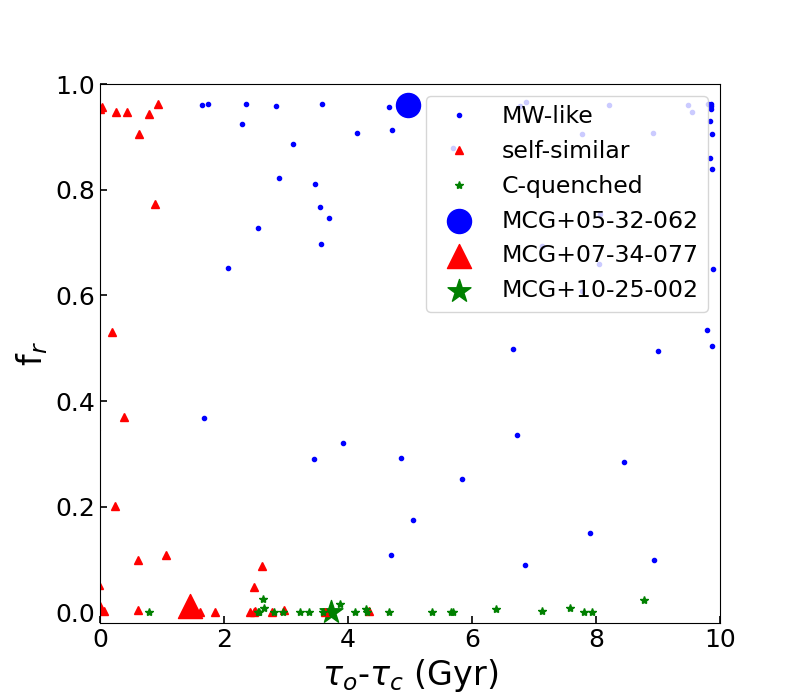}
     \caption{Distributions of difference of the gas infall time between the outer and inner region of the galaxy, against the strength of the radial inflow for our sample galaxies. MWAs behave like Milky Way are shown with blue dots, while the self-similar galaxies are shown in red triangles and centrally quenched galaxies in green stars. The larger symbols
     correspond to the three example galaxies with different types shown in Figures~\ref{fig:example_MWA}, \ref{fig:example_ss}, and~\ref{fig:example_cq}.}    
     \label{fig:dis_para}
\end{figure}

In order to quantify this qualitative classification, in \autoref{fig:dis_para} we plot the value of the best fit mass inflow parameter as a function of the difference in timescales for star formation between the inner and outer parts of each galaxy.  As discussed above, the systems we have identified as self-similar display little by way of gradient in their gas infall timescales, while the centrally-quenched galaxies have very low wind parameters.  It is notable that MCG+05-32-062, and, by association, the Milky Way, is not an exceptional galaxy.  It displays a fairly strong wind-parameter coupling between its inner and outer parts, but one that is common to quite a few of the MWAs, and a difference in timescale for star formation between inner and outer parts that is about average.  Milky-Way-like galaxies really are quite like the Milky Way, albeit within a fairly wide variety of inferred evolutionary histories.

\subsection{Differences between the three types of MWAs}

\begin{figure*}
    \centering
    \includegraphics[width=0.95\textwidth]{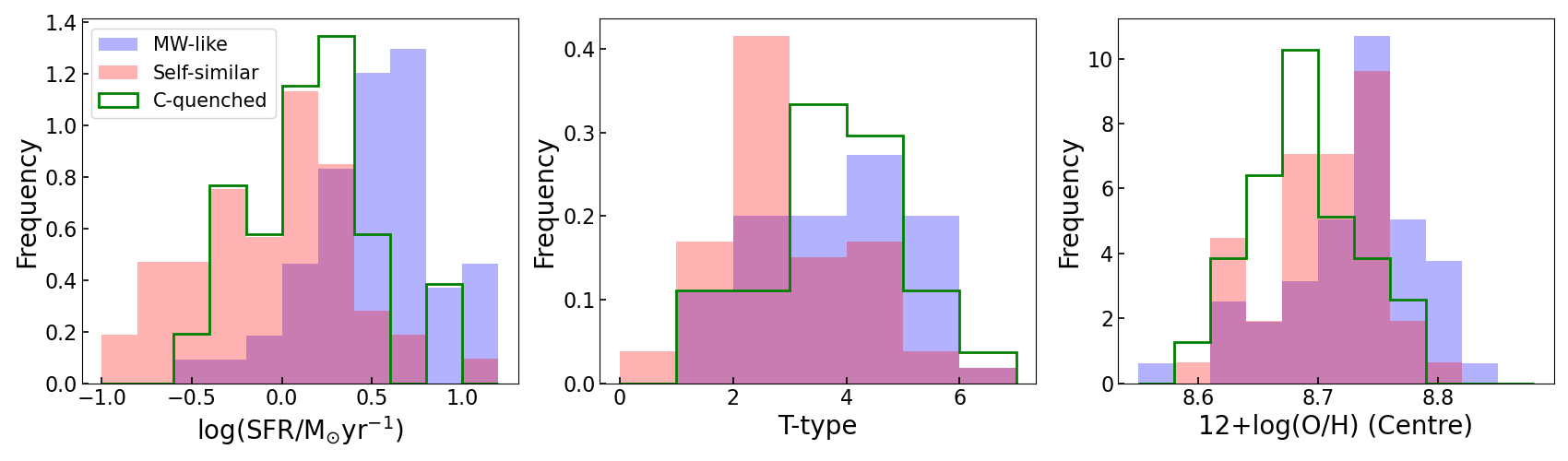}
     \caption{Distributions of the star formation rate (left), morphology T-type (middle), and gas phase metallicity (right) of the MWAs. Galaxies behave like the Milky Way are shown in blue, while the self-similar galaxies are shown in red and centrally quenched galaxies in green.
     }
     \label{fig:dis_SFR}
\end{figure*}

\begin{figure*}
    \centering
    \includegraphics[width=1.0\textwidth]{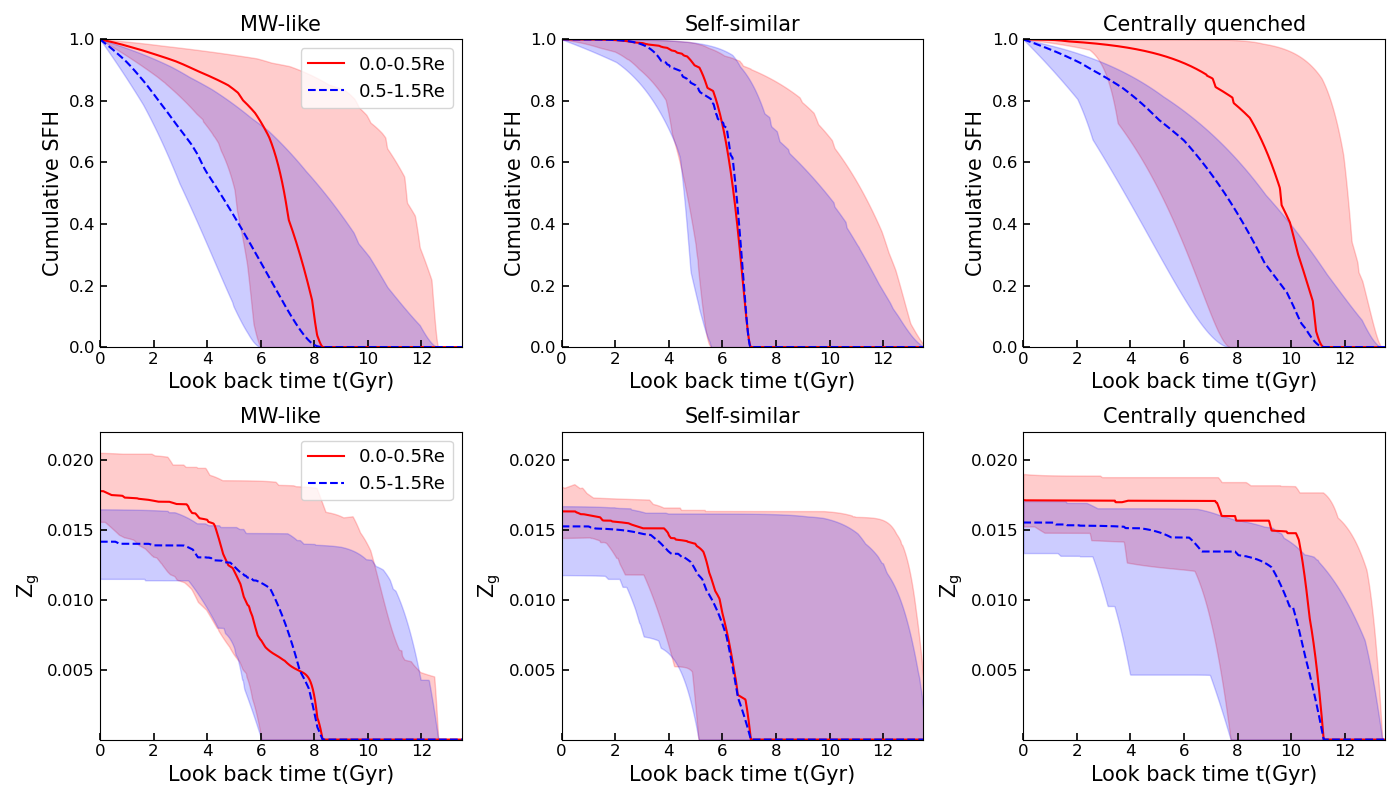}
     \caption{ The mass accumulation (top) and metallicity enrichment (bottom) histories for  MW-like (left), self-similar (middle) and centrally quenched (right) galaxies. In each panel, the evolution of the central and outer regions are shown in red and blue respectively. For each galaxy sample the lines correspond to the median values, with the shaded regions indicating the 16/84 percentiles of the distributions.
     }
     \label{fig:stat_SFH}
\end{figure*}

Having identified this diversity, a natural question is whether these differences manifest in any of the galaxies' other properties.  In \autoref{fig:dis_SFR} we present the distribution of various properties of the MWAs, divided into the three classes introduced above.  Although these galaxies were chosen to be a fairly homogeneous class of Milky-Way-like galaxies, it is clear that there are residual systematic differences in their properties that depend on their inferred evolutionary histories.

Of these differences, perhaps the least surprising is the dependence on the current star-formation rate (\autoref{fig:dis_SFR} left panel). The selection of MWAs deliberately avoided putting strong constraints on current SFR, beyond a related requirement of spiral morphology, so as not to select out galaxies that might be very similar to the Milky Way apart from very recent activity. Accordingly, there is quite a range in SFRs in the sample, and different types of galaxies show relatively narrow current SFR distributions with different typical values for each type: MW-like galaxies have the strongest recent star formation activities, with a median $\log({\rm SFR/M_{\odot}yr}^{-1})\sim0.5$, while the median values for self-similar and centrally quenched galaxies are $\log({\rm SFR/M_{\odot}yr}^{-1})\sim0.0$ and $\sim0.2$, respectively. In self-similar galaxies in which the evolution of inner and outer parts are close to identical, the timescale of star formation in the outer parts is somewhat dragged down to match the inner parts, leading to a systematically lower value for the current SFR.  Similarly, centrally-quenched galaxies seem to have completed more of their star-formation in the past, leading to a somewhat lower current rate than those more like the Milky Way.  A similar effect is seen in the T types of the galaxies (\autoref{fig:dis_SFR} centre panel): the median T-type values for MW-like, self-similar and centrally quenched galaxies are 4.0, 2.8, and 3.6, respectively. Such differences indicate that the suppression of star formation leads to systematically earlier Hubble types in the centrally-quenched and self-similar systems. Moreover, the right panel of \autoref{fig:dis_SFR} reveals a systematically lower gas-phase metallicity in the centre of centrally quenched galaxies compared to normal Milky-Way-like ones: the median gas-phase metallicities for MW-like, self-similar and centrally quenched galaxies are 12+log(O/H) $\sim 8.74$, 8.70, and 8.68, respectively. The lower central gas-phase metallicity of centrally quenched galaxies reflects the lack of metal-enriched gas inflow from the outer regions of these galaxies. In contrast, although not shown in the plot, there is no clear difference between the  gas-phase metallicities in the outer regions of different types of galaxies.

\begin{figure*}
    \centering
    \includegraphics[width=0.9\textwidth]{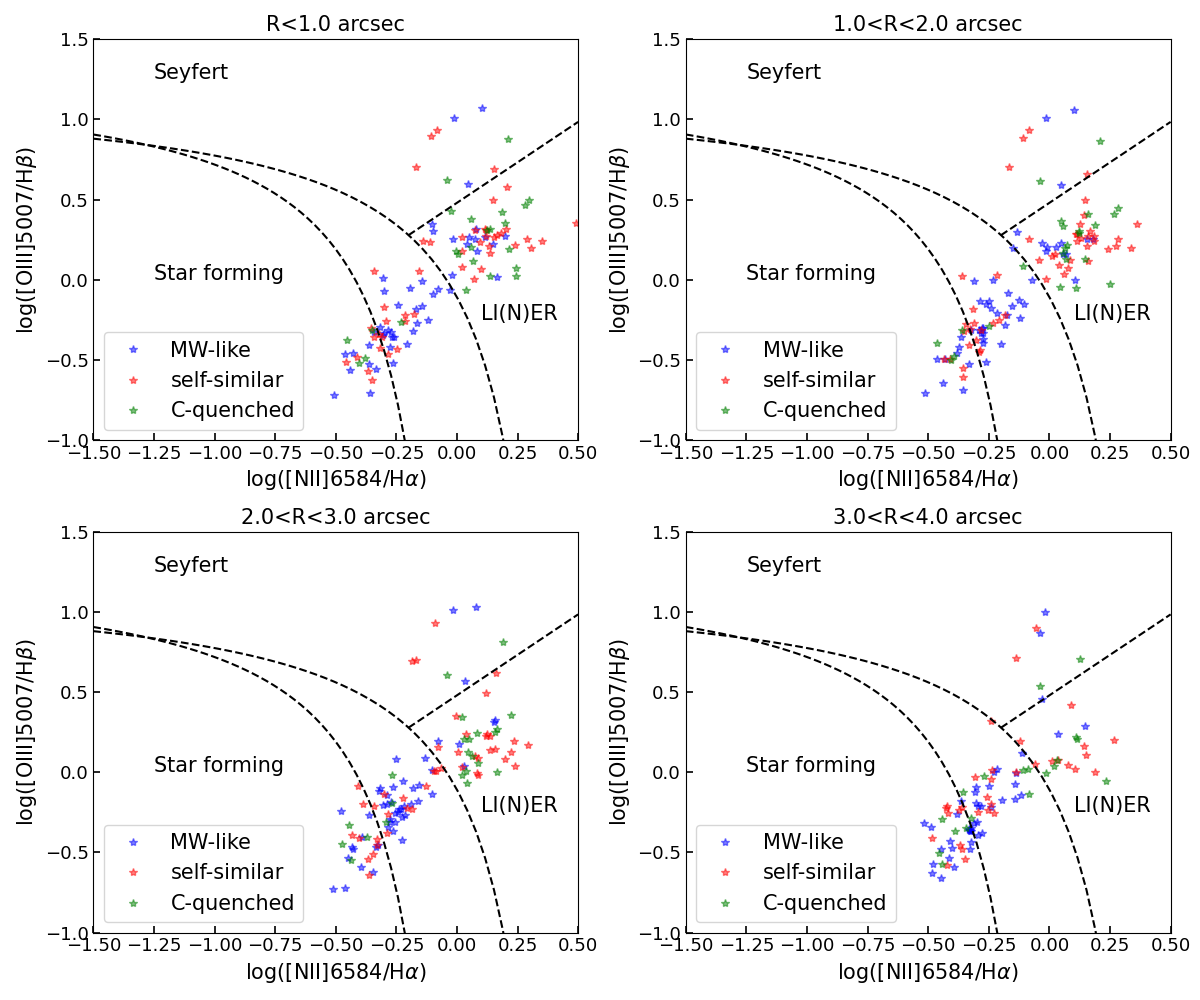}
     \caption{Distributions of our sample galaxies on the BPT diagram. From top-left to bottom-right, plots are for results from different positions of the galaxy, as labeled. In each plot, dashed lines are obtained from \citet{Kewley2001} and \citet{Kauffmann2003AGN} to divide the plane into regions of star forming, Seyfert, and LI(N)ER galaxies. Galaxies that behave like the Milky Way are shown as blue stars, while self-similar galaxies are shown in red and centrally quenched galaxies are shown in green.}
     \label{fig:dis_BPT}
\end{figure*}

Having found these present-day differences between different types of galaxies, we now turn to quantify the statistical differences in their inferred evolutionary histories. In \autoref{fig:stat_SFH} we investigate how the cumulative SFHs and ChEHs vary for the three different types of galaxies in our sample. Concerning the galaxies' SFHs, the top panels show that MW-like and centrally quenched galaxies exhibit clear differences in the mass accumulation histories between the central and outer regions, while in self-similar galaxies the two regions evolve almost identically. In addition,  MW-like galaxies have significant stellar mass formed in their central regions relatively late due to the radial flow in the recent 6~Gyrs, while the centres of almosr all centrally quenched galaxies finished their star formation 8~Gyrs ago. This echoes the evolution seen in the example galaxies shown in Figures~\ref{fig:example_MWA}, \ref{fig:example_ss}, and \ref{fig:example_cq}, with some scatter in the individual formation times and timescales. Concerning the galaxies' chemical evolution, self-similar and centrally quenched galaxies hold simple enrichment histories, while the centres of MW-like galaxies experience two major epochs of chemical enrichment.
As shown previously, this more complex behaviour is due to the metal-enriched gas brought in by the radial flow.
This metal-enriched radial flow also makes the centres of MW-like galaxies more metal-riched than the other types of MWAs, which is in line with the results inferred from the emission lines as shown in the right panel of \autoref{fig:dis_SFR}.

\subsection{Origin of the differences}
\label{subsec:origins}
These consistent differences between the three types of MWAs offer some assurance that our evolutionary model is capturing some real aspect of the formation and evolution of these galaxies, the remaining issue is to ask what might be driving such differences. Looking through the best fit parameters, we find the time at which galaxy formation began in this model varying systematically with those that are centrally quenched starting significantly earlier (as can be seen from the upper panels of \autoref{fig:stat_SFH}).  Perhaps these galaxies are the evolutionary successors of the Milky Way, which are further along in their lives and starting to see the inside-out quenching believed to occur in galaxies at these masses \citep[e.g.][]{Lin2019}.

One might suspect that the environment plays a role.  While in Milky-Way mass galaxies the effect of the environment may be relatively weak \cite[e.g.][]{Peng2012}, 
some simulation work \citep[e.g.][]{Garrison-Kimmel2019,Santistevan2020} suggests that Milky-Way-like galaxies in pairs -- such as the MW/M31 pair --  
typically formed their stars earlier than galaxies in isolated environments. 
Our sample does contain~8 galaxies with relatively massive nearby companions ($M_* >10^{10}\,{\rm M}_{\odot}$). The median gas infall time of these eight galaxies is~8.8\,Gyrs ago, while for the rest of the sample it is~7.4\,Gyrs ago. 
The Milky Way itself formed most of its stars around 10\,Gyrs ago, which is 2\,Gyrs earlier than the average value of the MW-like galaxies (see the top-left panel of \autoref{fig:stat_SFH}). These results, notwithstanding the small sample size, do support the simulation results. However, since environmental effects tend to change mainly the global properties of galaxies \citep[e.g.][]{Zheng_etal2019}, we would not necessarily expect to see any significant environmental imprint on the radial gradients of the stellar population properties we study. Among the 8 galaxies with massive companions, we find 4 MW-like, 3 self-similar and 1 centrally quenched galaxy, a distribution roughly similar to the entire sample.  We have also investigated a range of other environmental indicators, such as local density and cluster/group membership, and find no correlation with the evolutionary history of the MWAs. We thus conclude that the environment -- such as belonging to a massive pair -- may affect some the galaxies' global properties such as their formation times, but it is not likely to affect the internal secular evolution that we have studied here.

Another candidate sometimes invoked to try to explain evolutionary differences is feedback from AGN activity, which might, for example, cause an early shut-down of star formation (see \citealt{King2015} for a review), particularly in its vicinity. To investigate whether there is any current signature of AGN activity in the various types of MWA, \autoref{fig:dis_BPT} shows the BPT diagram, of emission line ratios log([OIII]5007/H$\beta$) versus log([NII]6584/H$\alpha$) that is often used to identify AGN activity, for different radii in the galaxies in this sample, colour coded by the evolutionary types identified above. It is apparent that there are intriguing systematic differences, with the self-similar galaxies tending to lie in the LINER part of the diagram associated with modest AGN activity; to a lesser extent, we see the same thing in the centrally-quenched galaxies.  However, if we look away from the centres of the galaxies, working toward the right in the panels of \autoref{fig:dis_BPT}, we see that although the galaxies all tend to move toward the star-forming region in the lower left of the BPT diagram, this effect persists, suggesting that the source of ionisation is not just an AGN.  Such extended LI(N)ER-like emission has been found in previous studies using spatially resolved spectroscopy \citep[e.g.][]{Belfiore2016}, but its ionizing source is still unclear. It is possible that hot post-AGB stars may be responsible for such extended ionisation \citep{Yan2012}, but there is also evidence that the central AGN emission can affect regions significantly larger than the spatial resolution of MaNGA \citep{Chen2019}.  There is therefore at least a hint that the central quenching of some galaxies could be related to AGN activity.

\subsection{Comparison with other studies}

Our work is the first one to explore the detailed spatially resolved chemical evolution of galaxies beyond the local group, where observations of individual stars are not available. It is therefore not possible to make direct like-with-like comparisons with previous studies. Nevertheless, there is some related work that is worth discussing in the context of our results.

In order to validate our modellining and fitting procedures, and to draw physical conclusions, we select MWAs that potentially have similar evolutionary histories to the Milky Way, and use the resolved stellar population properties of the MW
itself as a reference. There have been a plethora of Galatic chemical evolution models aimed at understanding the chemical content and evolution of the MW. The Galaxy's resolved stellar populations allow very detailed chemical evolution modelling (see reviews by \citealt{Matteucci2021}). For example, the classical two-infall model by \cite{Chiappini1997}, which assumes two distinct infall episodes, has now been widely used in many chemical evolution studies of the Milky Way, with refinements on various aspects \citep[e.g.][]{Grisoni2017,cote2017,Rybizki2017,Prantzos2018,Spitoni2019,Spitoni2021}. Another commonly cited model \citep{Schonrich2009} includes prescriptions for the radial mixing of gas and stars, and is expanded by a number of follow-up works \citep[e.g.][]{Marcon-Uchida2010,Minchev2013,Fu2013,Andrews2017}. Compared with these studies, our chemical evolution model has simpler assumptions; for example, only one epoch of gas infall is considered, and the radial migration of stars has not been included. However, as shown through this paper, our model is remarkably successful at describing simultaneously the mass accumulation and chemical enrichment of both the inner and outer regions of the Galaxy, as well as the bimodality distribution of stars on the [O/Fe] vs. [Fe/H] plane. 
Our predicted rate and speed of the radial flow are also roughly consistent with published values \citep[e.g.][]{Lacey1985,Portinari2000,Spitoni2013,Vincenzo2020}. These results indicate that our simplified models are able to catch the main physics governing the evolution of Milky-Way-like galaxies.

A key difference between our models and those aimed at understanding the MW is that we are investigating external galaxies. Although chosen to be MWAs, these galaxies are expected to have followed somewhat different evolutionary paths. Even within MW-like sample that we have identified, we detect some variations in their detailed SFHs, CheHs (\autoref{fig:stat_SFH}), and the physical parameters that govern their evolution, such as the strength of their radial flows. 
These differences prevent us from adopting models that are too specific, and very detailed comparisons between the derived physical parameters with those obtained from the Milky Way are not justified. 
We are interested in global trends concerning the evolution of the galaxies, rather than delving too deeply into the exact values of the derived model parameters. We find that we have reached reasonable consistency between the results of our models and those from previous work. Furthermore, our current understanding of the evolution of the Milky Way can broadly apply to other Milky-Way-like galaxies, even though significant differences in the galaxies' histories are found.

There is also a number of studies of MWAs in the literature that focus on different aspects of their stellar populations and other properties. For example, the work of \cite{Fraser-McKelvie2019} investigated the SFRs of MWAs from both SED fits and mid-infrared calibrations. They found that the Milky Way's SFR falls within $2\sigma$ of the MWA distribution, which is in line with our results for the star formation activity of our sample galaxies (\autoref{fig:dis_SFR}). \cite{Boardman2020} analysed MWAs in terms of their kinematics and stellar populations, as well as their 
ionized gas contents. Focusing on the gradients of these populations, they found significant variations among different MWAs and the Milk Way. Our work on MWAs is limited to comparing the inner and the outer regions, and thus doesn't provide detailed gradient measurements. Nevertheless, we do find that the variations in the evolution of MWAs correlate deeply with the gradients of their stellar populations -- galaxies with the strongest gradient in stellar ages are generally centrally quenched, while self-similar galaxies have almost no radial gradient in their physical properties. \cite{Krishnarao2020} investigated the effect of bars in MWAs and found that barred galaxies show stronger suppression of star formation and an increase in LI(N)ER-like spectra in their inner regions. We did not find that the presence of a bar affects the probability of a galaxy being placed in any of the three categories. However, as our division of the inner/outer regions does not take into account the possible existence of a bar, and given that the size of our sample doesn't allow for robust statistical analysis of the bar properties, we cannot rule out that bars may play some role in shaping the evolution of MWAs. 
We plan to explore this issue in the future with larger galaxy samples.

\section{Summary}
In this paper, we investigate the formation and evolution of a sample of 138 objects selected from the SDSS-IV/MaNGA survey to be analogues of the Milky Way, in order to assess how typical our galaxy is, and how much variety one sees in the evolutionary history of such an apparently homogeneous group of objects. To facilitate a comparison between these MWAs and the Milky Way, we have fitted their spectra with a simple self-consistent model that tracks the star formation and chemical evolution of the inner and outer parts of each target.  The model encompasses the core physics of star formation, gas infall and outflow, and a radial gas flow that connects the two regions. In simultaneously fitting the spectra from the inner and outer regions, we use both their absorption lines and auxiliary information from emission lines. The resulting best-fit model gives a picture of the star-formation and chemical evolution in both the inner and outer parts of each galaxy.  This information is then processed to predict the evolution of $\alpha$ element abundance in these galaxies; with this analysis, we can predict how the age -- metallicity -- alpha enhancement parameter space is populated by stars in both the inner and outer parts of each galaxy, which can be compared to the equivalent information obtained directly from individual stars in the Milky Way.  The main results of this comparison are as follows:
\begin{itemize}

\item
There are good number of MWAs with star formation and chemical histories very similar to our Milky Way. Initial star formation in the inner regions of these galaxies typically occurs on a shorter timescale than in their outer parts, but low-level star formation continues in the inner region, driven largely by processed material that flows in from the outer parts, helping to build up their radial metallicity gradients.

\item
This process also naturally builds up a bimodality in the stellar population in the parameter space of [O/Fe] versus [Fe/H], similar to that seen in the Milky Way: two distinct populations, one high in [O/Fe] but low in [Fe/H] and the other with low [O/Fe] but high [Fe/H], are predicted throughout these galaxies. In addition, this bimodality has a radial variation like that in the Milky Way, such that the two populations are both strongly populated in the central regions of galaxies, while the outer regions are dominated by the low $\alpha$ population. From the chemical evolution model, we learn that this arrangement can arise because the outer regions have a long timescale of star formation, which allows SNIa explosions to pollute the ISM with enough iron, leading to the dominance of the low-$\alpha$ population at large radii. At small radii, the high-$\alpha$ population is formed more rapidly at the beginning of the galaxy's evolution, while the low-$\alpha$ population formed subsequently from processed gas that flows in  from the outer regions.

\item 
While many MWAs seem to follow this Milky-Way-like pattern, others do not.  Some show a self-similar structure, with comparable timescales for star formation at all radii, and accordingly little by way of gradients in their stellar properties.  Others are centrally-quenched galaxies in which stars in the central regions formed rapidly, with no significant radial inflows and hence no late-time star formation in their inner regions.

\item
Comparing the detailed properties of these galaxies, we find that, despite being selected for their homogeneity, there are residual differences in the star-formation rates and Hubble types of MWAs with these different inferred evolutionary pasts.  In addition to providing clues as to the reasons for their differing pasts, perhaps implying that some are further along an evolutionary sequence than others, these differences give some confidence that the histories that we have been able to derive from the spectra are based in reality.  In looking for causes of these differences, we could find no environmental dependence, but intriguing clues that AGN activity could have suppressed the inflow of gas in some galaxies.

\end{itemize}

Semi-analytic spectral fitting has provided us with a powerful new tool for archaeologically exploring the chemical and star-formation evolution of galaxies in a self-consistent manner.  As this paper has shown, the methodology can readily be extended to incorporate additional processes, such as the coupling between different regions within a single galaxy, and extra physics like the time delay in recycling iron from Type Ia supernovae.  With these enhancements, we can model quite subtle properties of a galaxy's stellar population, such as their distribution in the age -- metallicity -- alpha enhancement space, for direct comparison with detailed data from the Milky Way.  From such a comparison, we learn that, while not all Milky-Way-like galaxies are like the Milky Way, many are.

\section*{Acknowledgements}
SZ, AAS and MRM acknowledge financial support from the UK Science and Technology Facilities Council (STFC; grant ref: ST/T000171/1).

For the purpose of open access, the authors have applied a creative commons attribution (CC BY) to any journal-accepted manuscript.

Funding for the Sloan Digital Sky Survey IV has been provided by the Alfred P. 
Sloan Foundation, the U.S. Department of Energy Office of Science, and the Participating Institutions. 
SDSS-IV acknowledges support and resources from the Center for High-Performance Computing at 
the University of Utah. The SDSS web site is www.sdss.org.

SDSS-IV is managed by the Astrophysical Research Consortium for the Participating Institutions of the SDSS Collaboration including the Brazilian Participation Group, the Carnegie Institution for Science, Carnegie Mellon University, the Chilean Participation Group, the French Participation Group, Harvard-Smithsonian Center for Astrophysics, Instituto de Astrof\'isica de Canarias, The Johns Hopkins University, Kavli Institute for the Physics and Mathematics of the Universe (IPMU) / University of Tokyo, Lawrence Berkeley National Laboratory, Leibniz Institut f\"ur Astrophysik Potsdam (AIP), Max-Planck-Institut f\"ur Astronomie (MPIA Heidelberg), Max-Planck-Institut f\"ur Astrophysik (MPA Garching), Max-Planck-Institut f\"ur Extraterrestrische Physik (MPE), National Astronomical Observatories of China, New Mexico State University, New York University, University of Notre Dame, Observat\'ario Nacional / MCTI, The Ohio State University, Pennsylvania State University, Shanghai Astronomical Observatory, United Kingdom Participation Group, Universidad Nacional Aut\'onoma de M\'exico, University of Arizona, University of Colorado Boulder, University of Oxford, University of Portsmouth, University of Utah, University of Virginia, University of Washington, University of Wisconsin, Vanderbilt University, and Yale University.

\section*{Data availability}
The data underlying this article were accessed from: SDSS DR17 \url{https://www.sdss.org/dr17/manga/}. The derived data generated in this research will be shared on request to the corresponding author.

\label{sec:summary}
%%%%%%%%%%%%%%%%%%%%%%%%%%%%%%%%%%%%%%%%%%%%%%%%%%
\bibliographystyle{mnras}
\bibliography{szhou} % if your bibtex file is called example.bib

% Don't change these lines
\bsp	% typesetting comment
\label{lastpage}
\end{document}